\documentclass[twocolumn,preprintnumbers,superscriptaddress,nofootinbib,aps,prd,floatfix,letterpaper]{revtex4-2}

\usepackage{amsmath}
\usepackage{amssymb,dsfont}
\usepackage{bm}
\usepackage{booktabs}
\usepackage{enumerate}
\usepackage[shortlabels]{enumitem}
\usepackage{graphicx}
\usepackage{hyperref}
\usepackage{multirow}
\usepackage{xcolor}

\hypersetup{colorlinks, linkcolor=black, urlcolor={blue}, citecolor={blue}}

\definecolor{grn}{HTML}{2c9119}

\begin{document}

\title{Equivariant Energy Flow Networks for jet tagging}

\author{Matthew J.\ Dolan}
\email{matthew.dolan@unimelb.edu.au}
\affiliation{ARC Centre of Excellence for Dark Matter Particle Physics, School of Physics, The University of Melbourne, Victoria 3010, Australia}
\author{Ayodele Ore}
\email{ayodeleo@student.unimelb.edu.au}
\affiliation{ARC Centre of Excellence for Dark Matter Particle Physics, School of Physics, The University of Melbourne, Victoria 3010, Australia}

\begin{abstract}
    Jet tagging techniques that make use of deep learning show great potential for improving physics analyses at colliders. One such method is the Energy Flow Network (EFN) --- a recently introduced neural network architecture that represents jets as permutation-invariant sets of particle momenta while maintaining infrared and collinear safety. We develop a variant of the Energy Flow Network architecture based on the Deep Sets formalism, incorporating permutation-equivariant layers. We derive conditions under which infrared and collinear safety can be maintained, and study the performance of these networks on the canonical example of W-boson tagging. We find that equivariant Energy Flow Networks have similar performance to Particle Flow Networks, which are superior to standard EFNs. \textcolor{black}{Due to convergence and generalisation issues, the same improvement was not observed when extending Particle Flow Networks with equivariant layers}. Finally, we study how equivariant networks sculpt the jet mass and provide some initial results on decorrelation using planing.
\end{abstract}

\maketitle

% Section 1
\section{Introduction}

In the past half-decade, there has been a burst of activity surrounding the 
application of deep learning methods (particularly neural networks) to a 
variety of problems in collider physics. Much of this research has been 
focused on tagging jets \cite{Lonnblad:1990qp, Cheng:2017rdo, Luo:2017ncs, 
efn, Kasieczka:2018lwf, Qu:2019gqs, calo, jet-images, dl-colour, baldi, 
unimelb, Pearkes:2017, Almeida:2015jua, efp, spectra, message-pass,jedi, 
qcd-aware, lund, QCDorWhat, autoencoder, Kasieczka:2017nvn, 
Macaluso:2018tck, Butter:2017cot,MUST}. Jets are collimated sprays 
of particles produced from energetic partons and contribute significantly 
to the complexity of a collider event. Optimal methods of jet tagging are 
therefore highly desirable in order to maximally exploit the physics 
capabilities of the LHC and future colliders. Examples include  quark/gluon 
discrimination \cite{qgtag, qgdiscrim, FerreiradeLima:2016gcz, Frye:2017yrw, 
Davighi:2017hok, Komiske:2018vkc, Larkoski:2019nwj} and the discrimination 
of QCD jets from overlapping jets produced by the hadronic decay of a 
boosted heavy resonance~\cite{butter, Salam:2009jx, Plehn_2012, 
Dasgupta:2015yua, nsub,Ju:2020tbo}, a topic of much interest at the LHC. 
Indeed, the ``standard candle" for jet tagging studies is the 
identification of boosted W-bosons from their decay products, a process 
which we study in this work.

The new ML-based methods leverage the fine resolution of modern detectors 
through their ability to automatically construct powerful discrimination 
variables from low-level inputs. By choosing a suitable representation for 
the inputs, existing deep learning models can be directly applied to the 
task of classifying jets. For example, multilayer perceptrons (MLPs) can be 
trained on lists of jets' constituent particle momenta~\cite{Pearkes:2017}, 
and convolutional neural networks (CNNs) can be trained on jet images~
\cite{calo, jet-images, unimelb, dl-colour, baldi}. These methods have led 
to substantial improvements on the performances of traditional taggers. 
However, the jet representations they use have  drawbacks. For example, 
jet images tend to be sparsely populated by particles, making a large 
number of the input pixels redundant. On the other hand, some methods based 
on the treatment of jets as lists (such as recurrent neural networks) fail 
to account for the fact that the particles in a jet have no intrinsic 
ordering. Accordingly, it is important to continue to thoroughly explore 
the landscape of machine learning-based taggers, and better understand 
their performance through analytic calculations~\cite{Kasieczka:2020nyd}.

A specific recent example is the Energy Flow Network (EFN), introduced in 
Ref.\,\cite{efn}. EFNs are based on the Deep Sets framework~\cite{deep-sets}, 
which treats data as point clouds -- sets of points in space. Point clouds do 
not have a fixed length or notion of ordering, similar to particle jets at 
colliders. The Deep Sets framework demonstrates how to incorporate permutation 
invariance and equivariance of the data -- jet constituents in our case. The 
permutation-invariant case was extensively studied in Ref.\,\cite{efn} which 
treats jets as sets of particle four-momenta, and  was also able to show that 
the resulting observables can be made infrared and collinear (IRC) safe. More 
general non-IRC safe networks were also studied, and dubbed Particle Flow 
Networks (PFNs).

Along with permutation-invariant networks the Deep Sets framework can 
incorporate permutation equivariance. A permutation-equivariant function is 
one which commutes with the permutation operation. That is, for input data 
$\vec{x}$, a permutation $\pi$ and a function $f$ equivariance requires 
$f(\pi(\vec{x}))=\pi(f(\vec{x}))$. Clearly this requires the dimensionality 
of the output of $f$ to be the same as the input. For binary classifiers the 
output is usually one-dimensional. Accordingly, it is straightforward to have 
an entire network which exhibits invariance of its output under permutation of 
the input parameters, but equivariance is implemented on individual internal 
layers of the network. While equivariance under \textcolor{black}{transformations of inputs } is less obviously motivated than invariance, \textcolor{black}{it has proven to benefit performance on some tasks. Most notably, the convolutions performed within a CNN are translation-equivariant\,\cite{LeCun} and are responsible for such models' suitability to images, which are naturally symmetric to translations. Equivariance to more complex symmetries has also been investigated such as in Refs.\,\cite{group-equi, Thomas2018TensorFN, Bogatskiy:2020tje}. Ref.\,\cite{Bogatskiy:2020tje} constructs networks that are equivariant under 
the action of the Lorentz group.}
Since the Lorentz group is a Lie group and hence 
continuous, this necessitates a different approach from the finite permutation groups 
relevant to this work.

In this paper, we continue the study of Energy Flow Networks by implementing 
permutation-equivariant network layers into the EFN architecture. We are able 
to find a variant which maintains infrared and collinear safety of the observables 
learned by the resulting network. We study the performance of equivariant Energy 
Flow and Particle Flow networks on boosted $W$-tagging, including some networks 
that use information about the jet constituent ID. While the equivariant EFNs slightly 
outperform the invariant EFNs, we find that this can change under decorrelation of the 
network output using planing. We also consider some possible future directions.

% Section 2
\section{Energy Flow Networks}

\begin{figure*}[tp!]
    \centering
    \includegraphics[width=0.92\textwidth]{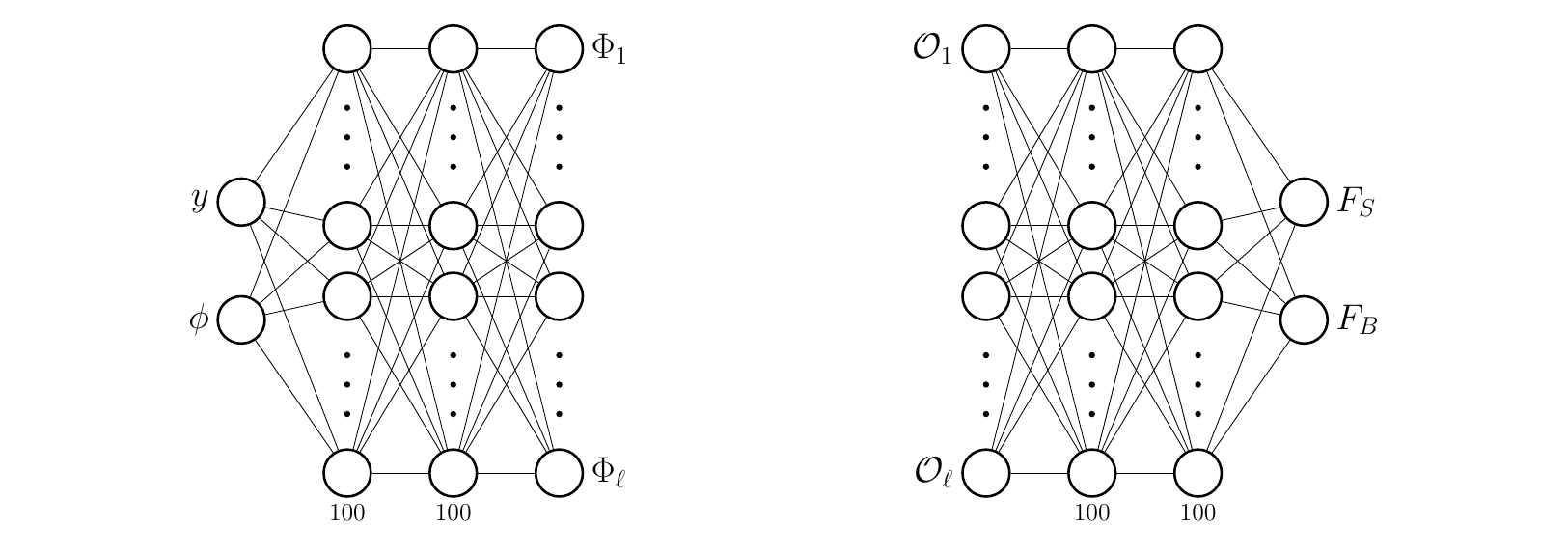}
    \caption{Examples of the two multilayer perceptrons comprising an EFN. The per-particle mapping (left) produces $\ell$ filters and the function $F$ (right) takes the observables $\mathcal{O}_a = \sum_i z_i\Phi_a(y_i, \phi_i)$ to a probability that the input was a signal jet ($F_S$), or a background jet ($F_B$).}
    \label{fig:arch}
\end{figure*}

In principle, the output of any neural network that takes jets or jet constituents as 
input can be interpreted as a jet observable. This output could be sensitive to the 
ordering of the constituents, contrary to the notion that collider events can be viewed 
as unordered list of particles. This has motivated the development and application of 
neural network architectures which are insensitive to the ordering of the jet constituents, 
where jets are viewed as point clouds: a set of data points in space. The representation of 
functions on sets from which neural network architectures may be modeled has been much 
studied recently in the ML community \cite{deep-sets, pointnet, bruno-perm, janossy-perm,
 adversarial-perm}. Collider physics studies in this vein include the use of architectures 
 based on Graph Neural Networks such as Refs.\,\cite{Qu:2019gqs} and related 
 Message Passing Neural Networks~\cite{message-pass,Ren:2019xhp,Abdughani:2020xfo}. 

We will explore the Energy Flow Network (EFN), an example  which is based closely on the Deep 
Sets model introduced in Ref.\,\cite{deep-sets}. Among deep learning models for jet tagging 
EFNs have the important feature that they are engineered to enforce infrared and collinear 
(IRC) safety on the network output. IRC safety is the requirement that a jet observable be 
invariant under soft and collinear emissions. We will demonstrate IRC safety of equivariant 
EFNs in Section~\ref{sec:augmented}.

\subsection{Deep Sets framework}
\label{deepsets}

Deep Sets~\cite{deep-sets} provides a framework for constructing neural networks 
that incorporate permutation invariance of the inputs. An Energy Flow Network is 
an implementation of this model which acts on jets (considering jets as sets of 
particles with associated momenta), with the additional requirement of IRC safety. 

The Deep Sets framework represents permutation-invariant functions $f$ in the 
following way. For a function acting on a set \ensuremath{X}, permutation 
invariance is ensured by first mapping the individual elements of \ensuremath{X} 
to an intermediate \emph{latent space} using a function $\Phi$. The results 
$\Phi(x_i)$ are then summed in the latent space. Finally, a different map $F$ takes 
the result of the summation in the latent space to the desired range of $f(X)$. Given 
that the sum is invariant under permutations of elements in \ensuremath{X}, the same 
property is inherited by the total function $f$. The sum also allows such networks 
to accept inputs of varying length.

The authors of Ref.\,\cite{deep-sets} prove the following
\begin{center}
% \begin{minipage}{1.0\textwidth}
\begin{minipage}{\linewidth}
\textbf{Deep Sets Theorem:} \textit{A function $f$ operating on a set \ensuremath{X} having 
elements from a countable universe $\mathfrak{X}$ is invariant to the permutation of 
instances in \ensuremath{X} if and only if it can be decomposed in the form}
\begin{equation}
f(\ensuremath{X}) = F\left(\,\sum_{x\in \ensuremath{X}}\Phi(x)\right),
\end{equation}
\textit{for suitable transformations $\Phi$ and $F$.}
\end{minipage}
\end{center}

Ref.\,\cite{efn} applies this decomposition to jet observables, which we may consider 
as functions on the set of constituent particle features (four-momentum, charge, flavour etc.). 
Such observables can therefore be written in the form
\begin{equation}\label{eq:pfn-obs}
    \mathcal{O}\left(\left\{p_1,\:\cdots,p_M\right\}\right) = F\left(\sum_{i=1}^M\Phi(p_i)\right),
\end{equation}
where each $p_i\in\mathbb{R}^d$ is a vector containing $d$ features of particles in the jet, 
$\Phi:\mathbb{R}^d\to\mathbb{R}^\ell$ is a per-particle map to the latent space, and 
$F:\mathbb{R}^\ell\to\mathbb{R}$ is a continuous mapping. The index $i$ runs over the $M$ 
particles in the jet. In fact, even before applying the function $F$, each of the $\ell$ 
components of the summed per-particle maps is a jet observable. The final observable 
$\mathcal{O}$ is some combination of these, dictated by $F$. A number of familiar examples 
can be written in this way. For example,  the jet mass corresponds to choices $\Phi(p^\mu)=p^\mu$ 
and $F(x) = \sqrt{x^\mu x_\mu}$.

Ref.\,\cite{efn} makes use of the Stone-Weierstrass Approximation Theorem\:\cite{weierstrass} 
to derive the specialisation of Eq.\,\ref{eq:pfn-obs} to the case where $\mathcal{O}$ 
is IRC safe. They arrive at
\begin{equation}
    \label{irc-obs-decomp}
    \mathcal{O}(\left\{p_1,\;\cdots,p_M\right\})=F\left(\sum_{i=1}^Mz_i\Phi(\Hat{p}_i)\right),
\end{equation}
where the per-particle mapping $\Phi:\mathbb{R}^2\to\mathbb{R}^\ell$ now depends only on 
the angular coordinates of the jet constituent momenta through $\Hat{p}\equiv(y,\phi)$, 
and the sum is weighted by the transverse momentum fractions $z_i=p_{T,i}/\sum_jp_{T,j}$.

Finally, we comment on the dimensionality of the latent space. While any IRC safe observable 
can be represented this way, this may require a latent space of high dimensionality. In 
Ref.\,\cite{deep-sets}, the Deep Sets Theorem was only proven for the case where instances 
of the set \ensuremath{X} are elements of a countable domain. Ref.\,\cite{setfunclimit} argues 
that extending consideration to uncountable sets is necessary since continuity on countable set 
(such as $\mathbb{Q})$ is a vastly different condition to continuity on the (uncountable) reals. 
They show that the Deep Sets theorem can only be extended to \emph{finite subsets} of uncountable 
domains, with the condition that the latent dimension should be at least the number of elements 
in the set. That is, in order to ensure universal function approximation of a network in the 
Deep Sets framework, one should have $\ell\geq M$. This is consistent with the observation in 
Ref.\,\cite{efn} that an EFN's performance saturates near $\ell=64$, since the multiplicity 
distributions of the jet samples peak around 40--50. We will present results below for 
$\ell=128$ and 256.

\subsection{Architecture}
To realise Eq.\,\ref{irc-obs-decomp} as a neural network, Ref.\,\cite{efn} uses two 
multi-layer perceptrons (MLPs) to approximate the functions $\Phi$ and $F$. A schematic 
of this construction for an EFN with latent dimension $\ell$ is presented in Fig.\,\ref{fig:arch}. 
The network on the left represents the per-particle mapping and the number of nodes in its 
output is equal to the dimension of the latent space. Each of these output nodes is called 
a \emph{filter} and is a function of the jet constituents' angular coordinates $y$ and $\phi$. 
Each filter is transformed into a jet observable by performing the $p_T$-weighted sum 
$\mathcal{O}_a = \sum_iz_i\Phi_a(y_i,\phi_i)$, where the sum runs over particles in the 
input jet. This set of observables is then passed as input to the second network which 
represents the map $F$. As is customary for the application of neural networks to classification 
tasks, the output layer of $F$ has two nodes and uses the \emph{softmax} activation, which 
correlates these nodes such that their outputs take values in the interval $[0,1]$ and 
sum to unity. The output of the network may then be interpreted as probabilities, $F_S$ and 
$F_B$, that the input is a $W$ (signal) or QCD (background) jet respectively. However, since 
the outputs are correlated, the network really only represents a single IRC safe jet observable 
for which $W$ jets have a value close to 1, and QCD jets, 0 (or vice-versa).

% Section 3
\section{Augmented Energy Flow Network architecture}
\label{sec:augmented}

\begin{figure*}[t!]
    \centering
    \includegraphics[width=0.92\textwidth]{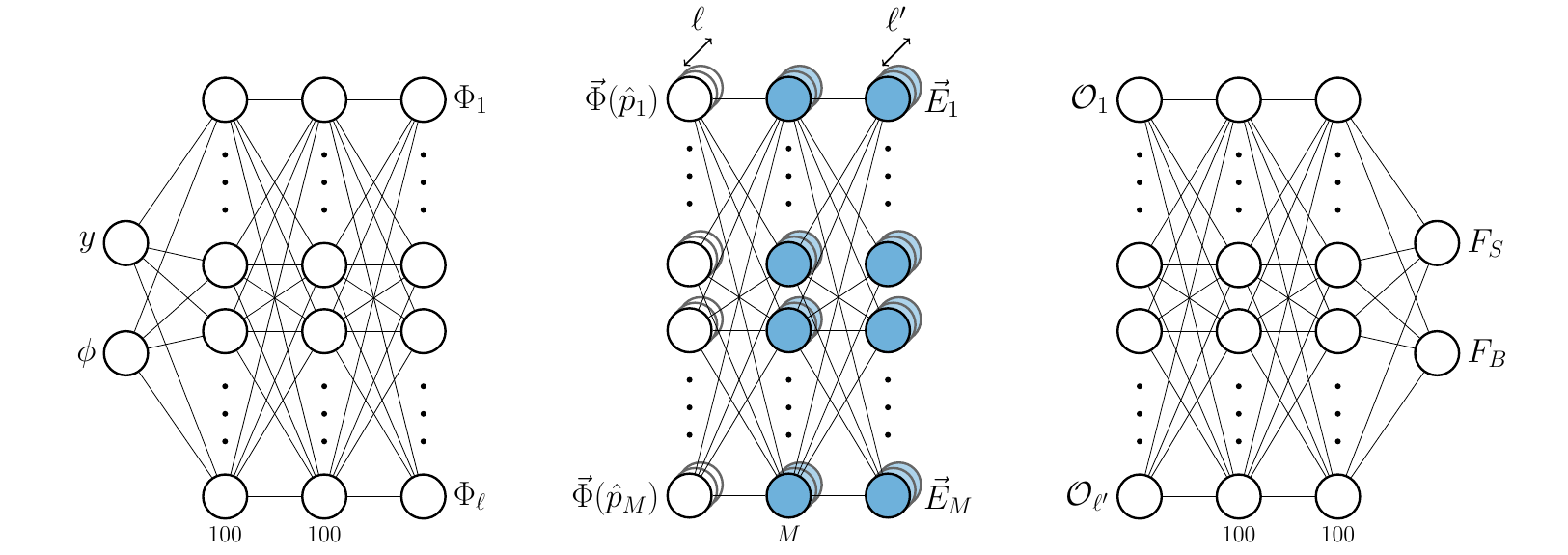}
    \caption{An example implementation of two equivariant layers (blue) into the EFN architecture. The first equivariant layer has $\ell$ input channels and the second has $\ell'$ output channels. A pooling operation is required to merge the vectors $E_i$ into a single vector $\Vec{\mathcal{O}}$ containing jet observables.}
    \label{fig:equi-arch}
\end{figure*}

In addition to presenting a permutation-invariant neural network architecture, on which the 
EFN is based, Ref.\,\cite{deep-sets} also defines permutation-\emph{equivariant} neural network 
layers. Equivariance of a function means that it commutes with permutations, rather than being 
invariant to them. Here we detail the way in which an equivariant layer can be incorporated into 
the existing structure of an EFN and seek an architecture that is able to maintain IRC safety of 
the network output.

\subsection{Equivariant network layers}\label{sec:layer-def}

Permutation equivariance constrains the form of neural network layers. Consider a network layer 
$E_\Theta(\bm{x})=\sigma(\Theta\bm{x})$, having activation $\sigma$, weight matrix 
$\Theta \in \mathbb{R}^{M\times M}$, and acting on a set $\bm{x}$ with elements $x_i\in\mathbb{R}$. 
A result from Ref.\,\cite{deep-sets}  states that such a layer  is permutation-equivariant if and 
only if $\Theta$ can be decomposed as
\begin{equation}
    \label{eqvsum}
    \Theta = \lambda\bm{I} + \gamma(\bm{1}\bm{1}^T), 
\end{equation}
with $\lambda, \gamma\in\mathbb{R}$, $\bm{I}$ =  $id\in\mathbb{R}^{M\times M}$ and 
$\bm{1}=(1,\:\cdots,1)^T\in\mathbb{R}^M$. That is, the matrix $\Theta$ has all   diagonal terms 
equal and all off-diagonal terms equal. The function $E$ therefore depends only on a linear 
combination of the input $\bm{x}$ itself and $\bm{1}\bm{1}^T\bm{x}$ -- a vector with each 
component being the sum of components of $\bm{x}$. The equivariance of the layer arises from 
the fact that this summation respects permutation symmetry. One can also consider functions $E$ 
whose arguments are not of the linear form $\Theta \bm{x}$. For example, the function
\begin{equation}
    \label{eqvmax}
    E(\bm{x}) = \sigma(\lambda\bm{I}\bm{x} + \gamma\,\text{max}(\bm{x})\bm{1}),
\end{equation}
also defines an equivariant network layer. Eqs~\ref{eqvsum} and \ref {eqvmax} are valid when the 
elements of the set $x_i\in\mathbb{R}$. However, the $x_i$ could themselves be vectorial 
$x_i\in \mathbb{R}^D$, so that $\bm{x}\in\mathbb{R}^{M\times D}$. In that case the parameters 
$\lambda$ and $\gamma$ can be promoted to matrices $\Lambda, \Gamma\in\mathbb{R}^{D\times D'}$ 
with $D'$ being the dimension of each output node. This leads to
\begin{equation}
    \label{eqvmatsum}
    E(\bm{x}) = \sigma(\bm{x}\Lambda + \bm{1}\bm{1}^T\bm{x}\Gamma),
\end{equation}
\begin{equation}
    \label{eqvmatmax}
    E(\bm{x}) = \sigma\big(\bm{x}\Lambda + \bm{1}\text{maxpool}(\bm{x})\Gamma\big),
\end{equation}
where maxpool is an operation that merges feature vectors in the set $\bm{x}\in\mathbb{R}^{M\times D}$ 
to a single summary vector in $\mathbb{R}^D$ with $i^{\rm{th}}$ component equal to 
$\max\left\{{x_j}^i\right\}_{j=1}^M$.

We will see that EFNs using these layers are IRC unsafe in general. In our studies below we 
will use layers of the form Eq.\,\ref{eqvmatmax} as an example of such a network, and further 
discuss IRC safety below.

\subsection{Layer implementation}

As well as introducing the equivariant layer itself, Ref.\,\cite{deep-sets} also discusses how 
such layers can be implemented in a specific model. Two properties of the equivariant layer will 
be exploited. The first is that the composition of equivariant layers is also equivariant. This 
can be seen by considering equivariant functions $f,g$ acting on a set $\bm{x}$ with $n$ elements 
and $\pi\in S_n$ a permutation. Then
{\setlength{\belowdisplayskip}{6pt}
\setlength{\abovedisplayskip}{6pt}
\begin{align}
    (f&\circ g)(\pi \bm{x})\notag\\&=f(g(\pi \bm{x})) = f(\pi g(\bm{x})) = \pi(f(g(\bm{x})) = \pi(f\circ g)(\bm{x}) \, ,
\end{align}}\noindent%
so the composition of $f$ and $g$ is equivariant. This implies that equivariant layers can be 
stacked to build deep models.

The second property is that following an equivariant layer with any commutative pooling operation 
(such as sum, mean or max) across set elements yields a permutation-\textit{invariant} function. 
This allows us to replace the sum over per-particle maps (the permutation-invariant function in 
the standard EFN architecture) with a sequence of equivariant layers followed by a pooling operation. 
The network is then able to learn its own particular permutation-invariant function by which to 
combine the per-particle information.

A schematic of the architecture is presented in Fig.\,\ref{fig:equi-arch}. Just as with the 
standard permutation-invariant EFN, on the left we have a dense network representing a per-particle 
map $\Phi:\mathbb{R}^2\to\mathbb{R}^\ell$ that acts on the $y$-$\phi$ plane. In the centre of the 
diagram the vector representation of each particle under this map is then passed as input to a 
number of equivariant layers shown with blue shaded nodes, the first of which has $\ell$ input 
\emph{channels}. The number of input (output) channels in an equivariant layer is defined as the 
dimensionality of each input (output) node. 

\textcolor{black}{It is important to note that the size of the equivariant layers must be at least the number of jet constituents in the input, which may vary during training. As such, our new architecture is not naturally able to accept variable-length sets. This is addressed by fixing the layer size, $M$, to be suitably large and zero-padding the inputs. Such treatment is not necessary for an EFN, since its architecture is completely independent of the input jet multiplicity. However, the number of parameters contained in an equivariant layer does not vary with $M$, so zero-padding inputs to this new architecture is not as troublesome as it is in some other examples.}

The final equivariant layer, having $\ell'$ output channels, is followed by a pooling operation 
that projects the $M\times \ell'$-dimensional output of the layer to a vector $\Vec{\mathcal{O}}$ 
of length\;$\ell'$. There is freedom to choose this operation, for example some of the models 
described in Sec.\,\ref{sec:models} use a maxpooling to construct the components of the vector 
according to $\mathcal{O}^a=\max\left\{{E_i}^a\right\}_{i=1}^M$, where ${E_i}^a$ is a component 
of an output node in the final equivariant layer, with $i$ labeling jet constituents and $a$ 
indexing the latent dimension. These $\mathcal{O}^a$ are permutation-invariant jet observables, 
in analogy with the $p_T$-weighted sum over filters in an EFN, and are passed as input to the MLP 
representing the function $F:\mathbb{R}^{\ell'}\to\mathbb{R}$.

\subsection{IRC safety}
\label{sec:irc-safety}
The networks generated from the method above are distinct from the IRC safe EFN studied in 
Ref.\,\cite{efn}, which corresponds to taking the pooling operation as a $z$-weighted sum and 
the function $E$ as the identity. As such, IRC safety of the jet observables from the new 
networks is not guaranteed. In this section, we aim to find an architecture that maintains 
IRC safety. This allows for a closer comparison with the EFNs of Ref.\,\cite{efn}.

There are a number of potential architectures of the type described in the previous section. 
We will consider an equivariant energy flow network denoted by EV-EFN, which for simplicity 
we assume has a just a single equivariant layer $E:\mathbb{R}^{M\times\ell}\to\mathbb{R}^{M\times\ell'}$. 
We will consider the implications of multiple equivariant layers later in the section. We 
may express the action of the network as
\begin{align}\label{evefn-action}
    \mathrm{EV\text{-}EFN}\,\left(\{p_i\}_{i=1}^M\right) = \big(F\circ P\circ E\big)(\textcolor{black}{\{\Phi(\Hat{p}_i)\}_{i=1}^M}),
\end{align}
where $P:\mathbb{R}^{M\times\ell'}\to\mathbb{R}^{\ell'}$
is the pooling operation, and $F:\mathbb{R}^\ell\to\mathbb{R}$ is the final function. To specify a network, 
we must make choices for the pooling operation $P$ and the form of the equivariant layer 
(using Eqs.\,\ref{eqvmatsum},\,\ref{eqvmatmax} above for instance etc.). The maps  $F$ and $\Phi$ 
are given by fully-connected MLPs. We can write the vector of jet observables $\Vec{\mathcal{O}}$ 
by dropping $F$ from Eq.\,\ref{evefn-action}. This gives
\begin{align}
    \Vec{\mathcal{O}} = \big(P\circ E\big)(\textcolor{black}{\{\Phi(\Hat{p}_i)\}_{i=1}^M}).
    \label{eq:obs}
\end{align}
Since we know that taking $P$ as a $z$-weighted sum ${P(\bm{x})=\sum_{i}z_ix_i}$ is an IRC safe 
choice in an EFN, we will only consider this operation for this section. This may not be the unique 
choice for ensuring IRC safety, but we were able to rule out some other possibilities. For example, 
a max function is not necessarily invariant to the collinear splitting of a particle.

In an EFN with equivariant layers, this choice of $P$ alone is not enough to ensure that the 
observables vector $\Vec{O}$ is IRC safe. The reason for this is that in the new architecture, 
the output nodes of the equivariant layer are not strictly per-particle maps. That is, the 
vector $E_i$ has dependence on the momentum of particles other than $i$. However, the 
fact that this mixing comes through an equivariant layer imposes helpful constraints. 

Specifically, permutation equivariance demands that the $i^{\rm{th}}$ output of the layer depends 
on inputs other than $i$ only through permutation-invariant functions of the inputs (in linear combination). 
For example, in Eq.\,\ref{eqvsum} for $\left(\Theta\bm{x}\right)_i$ the first term depends only 
on $x_i$ and the second term is proportional to the permutation-invariant sum $\sum_i x_i$. Thus, 
despite the outputs $E_i $ of the equivariant layer not being per-particle maps, they do have a 
unique dependence on the $i^{\rm{th}}$ particle.

In addition, the permutation-invariant functions that give the dependence on all particles can be 
identified as jet observables. We show below that the IRC safety of the $z$-weighted sum of such an 
output, which  depends on the linear combination of single particle information and a jet observable, 
is tied to the IRC safety of the jet observables in the equivariant layer. We therefore parameterise 
an equivariant layer as
\begin{equation}
    E(\textcolor{black}{\{\Phi(\Hat{p}_i)\}_{i=1}^M}) = \sigma\left(\Phi(\Hat{p}_i)\cdot\Lambda + \mathcal{Q}(\textcolor{black}{\{\Phi(\Hat{p}_i)\}_{i=1}^M})\cdot\Gamma\right),
\end{equation}
where $\mathcal{Q}\in\mathbb{R}^\ell$ holds jet observables that depend on the set of per-particle 
latent vectors, the matrices $\Lambda$ and $\Gamma$ are defined as in Eq.\,\ref{eqvmatsum}, and $\sigma$ 
is the activation function. \textcolor{black}{However, as it stands, the equivariant layer only takes
directional information $\Hat{p}$ as input through the filters $\Phi$. If $\mathcal{Q}$ is to be IRC 
safe, it must also receive the momentum fractions $z$ as input. Such a dependence can be added to the 
layer $E$ through $\mathcal{Q}$ while maintaining equivariance if $Q$ is understood to be permutation-invariant 
with regard to pairs $\left(\Phi(\Hat{p}_i), z_i\right)$. This yields}
\begin{equation}
    E(\textcolor{black}{\{\Phi(\Hat{p}_i)\}_{i=1}^M,z}) = \sigma\left(\Phi(\Hat{p}_i)\cdot\Lambda + \mathcal{Q}(\textcolor{black}{\{(\Phi(\Hat{p}_i),z_i)\}_{i=1}^M})\cdot\Gamma\right),
\end{equation}

Taking $P$ as a $z$-weighted sum and the discussion above into account, Eq.\,\ref{eq:obs} for a 
network with a single equivariant layer can be written out in more detail as 
\begin{equation}\label{eq:eqv-irc}
    \Vec{\mathcal{O}}=\sum_{i=1}^Mz_i\sigma\left(\Phi(\Hat{p}_i)\cdot\Lambda + \mathcal{Q}(\textcolor{black}{\{(\Phi(\Hat{p}_i),z_i)\}_{i=1}^M})\cdot\Gamma\right).
\end{equation}

We wish to show that the observable $\Vec{\mathcal{O}}$ is IRC safe if every component 
of $\mathcal{Q}$ is IRC safe. To study infra-red safety, we add an arbitrarily soft particle 
labeled by index $s$ that can have momentum in any direction:

\begin{align}
    \Vec{\mathcal{O}}_{IR} &= \lim_{z_s\to0}z_s\sigma\left(\Phi(\Hat{p}_s)\cdot\Lambda + \mathcal{Q}_{IR}\cdot\Gamma\right) \notag \\
    &\ \ \:+ \sum_{i=1}^Mz_i\sigma\left(\Phi(\Hat{p}_i)\cdot\Lambda + \mathcal{Q}_{IR}\cdot\Gamma\right)\notag\\
    &=\sum_{i=1}^Mz_i\sigma\left(\Phi(\Hat{p}_i)\cdot\Lambda + \mathcal{Q}_{IR}\cdot\Gamma\right) \, ,
    \label{eqv-ir-safety}
\end{align}
where \textcolor{black}{$\mathcal{Q}_{IR}=\mathcal{Q}(\{(\Phi(\Hat{p}_s),z_s)\}\cup\{(\Phi(\Hat{p}_i),z_i)\}_{i=1}^M)$}.

For collinear safety a particle splits with momentum fractions $\lambda z$ and 
$(1-\lambda)z$ with $\lambda\in [0,1]$. Without loss of generality we take this 
to be the first particle. Under a collinear splitting the observable becomes

\begin{align}
    \Vec{\mathcal{O}}_C &= \lambda z_1\sigma\left(\Phi(\Hat{p}_1)\cdot\Lambda + \mathcal{Q}_C\cdot\Gamma\right)\notag\\
    &\quad+ (1-\lambda)z_1\sigma\left(\Phi(\Hat{p}_1)\cdot\Lambda + \mathcal{Q}_C\cdot\Gamma\right)\notag\\
    &
    \quad
    + \sum_{i=2}^Mz_i\sigma\left(\Phi(\Hat{p}_i)\cdot\Lambda + \mathcal{Q}_C\cdot\Gamma\right)\notag\\
    &=\sum_{i=1}^Mz_i\sigma\left(\Phi(\Hat{p}_i)\cdot\Lambda + \mathcal{Q}_{C}\cdot\Gamma\right),
    \label{eqv-c-safety}
\end{align}
and \textcolor{black}{$\mathcal{Q}_{C}=\mathcal{Q}(\{(\Phi(\Hat{p}_1),\lambda z_1),(\Phi(\Hat{p}_1),(1-\lambda) z_1)\}\cup\{(\Phi(\Hat{p}_i),z_i)\}_{i=2}^M)$}.

If $\mathcal{Q}_{IR}=\mathcal{Q}_{C}=\mathcal{Q}$, then Eq.\,\ref{eq:eqv-irc} is recovered 
from both Eq.\,\ref{eqv-ir-safety} and Eq.\,\ref{eqv-c-safety}. Thus, $\Vec{\mathcal{O}}$ is 
IRC safe if $\mathcal{Q}$ is IRC safe. One obvious choice for the components $\mathcal{Q}^a$ is 
to again use the $z$-weighted sum over per-particle maps, $\mathcal{Q}^a=\sum_iz_i\Phi^a(\Hat{p}_i)$ \textcolor{black}{leading to}
\begin{equation}\label{eq:irc-equi-def}
    E(\textcolor{black}{\{\Phi(\Hat{p}_i)\}_{i=1}^M},z) = \sigma\left(\textcolor{black}{\{\Phi(\Hat{p}_i)\}_{i=1}^M}\cdot\Lambda + \sum_{j=1}^Mz_j\Phi(\Hat{p}_j)\cdot\Gamma\right)
\end{equation}
The IRC safe jet observables are then obtained by composing with the $z$-weighted sum. In full, they read

\begin{align}\label{1equiircobs}
    \Vec{\mathcal{O}} &= \sum_{i=1}^Mz_i{E_i}\big(\textcolor{black}{\{\Phi(\Hat{p}_i)\}_{i=1}^M},z\big) \notag\\
    &= \sum_{i=1}^Mz_i\sigma\left(\Phi(\Hat{p}_i)\cdot\Lambda + \sum_{j=1}^Mz_j\Phi(\Hat{p}_j)\cdot\Gamma\right).
\end{align}
Interestingly, one finds that \textcolor{black}{in certain cases the network is equivalent to a standard EFN. For example, replacing activations by the identity gives}
\begin{align}\label{eq:id-act}
    \Vec{\mathcal{O}} &= \sum_{i=1}^Mz_i\left(\Phi(\Hat{p}_i)\cdot\Lambda + \sum_{j=1}^Mz_j\Phi(\Hat{p}_j)\cdot\Gamma\right)\notag\\
    &=\sum_{i=1}^Mz_i\Phi(\Hat{p}_i)\cdot\Lambda + \sum_{j=1}^Mz_j\Phi(\Hat{p}_j)\cdot\Gamma\notag\\
    &=\sum_{i=1}^Mz_i\Phi(\Hat{p}_i)\cdot\left(\Lambda + \Gamma\right)\notag\\
    &\equiv\sum_{i=1}^Mz_i{\Psi}(\Hat{p}_i)\ ,
\end{align}
where the first step follows from the fact that the $p_T$ fractions sum to unity, and we 
have redefined the per-particle map $\Psi^a(\Hat{p}_i) \equiv \sum_{b=1}^\ell\Phi^b(\Hat{p}_i)\left(\Lambda^{ba} + \Gamma^{ba}\right)$. 
This is a consequence of the fact that IRC safety imposes a linear structure on the 
transformations prior to $\sigma$. Removing the activation allows the action of the 
equivariant layer to be absorbed into the definition of the per-particle map. \textcolor{black}{A similar result is reached even when including activations if one restricts the equivariant layer(s) with $\Gamma=0$ or $\Lambda=0$.  See the  Appendix\,\ref{app:equi-test} for further comment on this.}

Next, we must check that IRC safety is maintained in a network consisting of more 
than one such equivariant layer. Composing two layers leads to observables
\begin{align}
    \Vec{\mathcal{O}} &= \big(P\circ E\circ E'\big)(\textcolor{black}{\{\Phi(\Hat{p}_i)\}_{i=1}^M}).\notag\\
    &= \sum_{i=1}^Mz_i{E_i}\Big(E'\big(\textcolor{black}{\{\Phi(\Hat{p}_i)\}_{i=1}^M},z\big),z\Big)\notag\\
    &= \sum_{i=1}^Mz_i\sigma\Bigg(E'_i\big(\textcolor{black}{\{\Phi(\Hat{p}_i)\}_{i=1}^M},z\big)\cdot\Lambda \notag\\
    &\phantom{=\sum_{i=1}^Mz_i\sigma\Bigg(}+ \sum_{j=1}^Mz_jE'_j\big(\textcolor{black}{\{\Phi(\Hat{p}_i)\}_{i=1}^M},z\big)\cdot\Gamma\Bigg)\notag\\
    &=\sum_{i=1}^Mz_i\sigma\Bigg[\sigma\left(\Phi(\Hat{p}_i)\cdot\Lambda'+ \sum_{k=1}^Mz_k\Phi(\Hat{p}_k)\cdot\Gamma'\right)\cdot\Lambda\notag\\
    &\quad\;+\sum_{j=1}^Mz_j\sigma\left(\Phi(\Hat{p}_j)\cdot\Lambda'+ \sum_{k=1}^Mz_k\Phi(\Hat{p}_k)\cdot\Gamma'\right)\cdot\Gamma\Bigg]\ ,
    \label{2equiircobs}
\end{align}
where $E'$ denotes the first equivariant layer, which has matrix parameters 
$\Lambda'$ and $\Gamma'$. Thus the vector $\Vec{\mathcal{O}}$ is IRC safe. To see this, 
notice that the summand for each sum over particles is weighted by $z$ and that these 
weights constitute all of the $z$ dependence in the expression. The same treatment can 
be applied for additional stacked layers. Although it would be impractical to present 
the equations here, one finds that the sums over particles are always $z$-weighted, 
leading to IRC safety. This is because adding an additional equivariant layer corresponds 
to replacing each $\Phi(\Hat{p}_{\{i,j,k\}})$ by $E_{\{i,j,k\}}$. Since $E_{\{i,j,k\}}$ 
contains only IRC safe observables and $\Phi(\Hat{p}_{\{i,j,k\}})$, this replacement 
leaves the same $z$-weighted structure of the latent vectors.

% Section 4
\section{Training and performance}

Here we evaluate the performance of equivariant EFNs in discriminating jets from boosted 
$W$ bosons against the QCD background. We will consider both IRC safe and unsafe equivariant 
networks. The IRC unsafe networks will be contrasted with the performance of Particle Flow Networks 
(PFNs)~\cite{efn}. PFNs are similar to EFNs, but based around the general observable decomposition 
of Eq.\,\ref{eq:pfn-obs} and are thus IRC unsafe. We begin by detailing the data generation 
and preparation steps before outlining the construction of each model.

\begin{figure*}
    \centering
    \includegraphics[width=\textwidth]{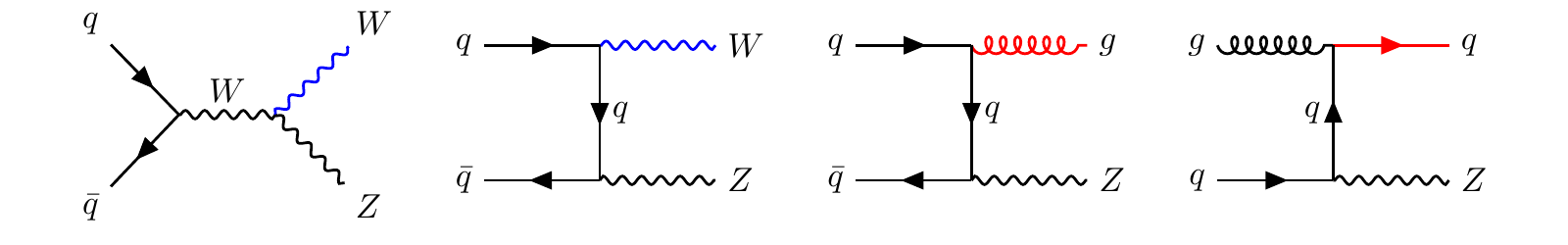}
    \caption{\footnotesize{Tree-level Feynman diagrams for the $W$ (blue) and QCD (red) contributions to the processes of interest. Rearrangement of the final diagram gives an additional $s$-channel contribution, $qg\to qZ$}.}
    \label{fig:feyn}
\end{figure*}
\subsection{Data generation and pre-processing}
The $W$ and QCD jets used for this work are obtained respectively from $pp\to Z+W$ and $pp\to Z + \text{jet}$ 
events generated and hadronised with \texttt{PYTHIA~8.2\textcolor{black}{26}}\:\cite{pythia} at centre of mass energy 
$\sqrt{s}=13\;\text{TeV}$. For the $W$ events, we use the \texttt{WeakDoubleBoson:ffbar2ZW} 
channel with the $W$ decaying to quarks. For the QCD events we use \texttt{WeakBosonAndParton:qg2gmZq} 
and \texttt{qqbar2gmZg} with the photon contribution switched off. The Feynman diagrams for the 
tree-level contributions to these processes are shown in Fig.\,\ref{fig:feyn}. In both cases, the 
$Z$ is required to decay invisibly to neutrinos. Multiple parton interactions as well as initial 
and final state radiation are left on.

To study network performance on jets of different transverse momenta we produce two datasets: 
the first containing jets with $p_T\in[250,300]$ GeV and the second with $p_T\in[500,550]$ GeV.
The datasets are produced as follows: after each event is generated, particles with $p_T$ less 
than $500$ MeV are removed with the intent of mitigating the influence of contamination from the 
underlying event. The event is then clustered by the anti-$k_t$ algorithm \cite{anti} using 
\texttt{FastJet~\textcolor{black}{3.3.3}}\,\cite{fastjet} with a jet radius of $R=1.0$ for the $p_T\in[250,300]$ GeV dataset 
and $R=0.6$ for the $p_T\in[500,550]$ GeV set.

If the hardest jet in the event has mass~$ m\in[65,105]$ GeV, rapidity $\left|y\right|<2.5$, and 
transverse momentum within the relevant range, the set of its constituent particle feature vectors 
is added to the dataset, with each vector in the format:  $(p_T,y,\phi,m,\text{\texttt{pdg\_id}})$. 
The field \texttt{pdg\_id} follows the Particle Data Group Monte Carlo particle numbering scheme \cite{pdg}. 
We centre the selection window slightly beyond the mass of the $W$ since the final jet will contain 
some additional radiation from the underlying event, translating the distribution. The final datasets 
contain 1 million $W$ and QCD jets in equal proportion and are split into three subsets for training 
(75\%), validation (15\%), and testing (10\%).

We employ four pre-processing steps intended to isolate the differences between the $W$ and QCD jets. 
Each jet in the dataset is first re-clustered using the $k_t$ algorithm\:\cite{genkt} with a jet radius 
of $R=0.2$ or $R=0.3$ for the high- or low-$p_T$ datasets respectively. The resulting subjets are ordered 
by $p_T$ and the following steps applied:

\begin{enumerate}
    \item \textbf{Normalise} the $p_T$ of all particles in the jet such that their sum is unity.
    \item \textbf{Translate} all particles in the $y$-$\phi$ plane such that the leading subjet lies at the origin.
    \item \textbf{Rotate} all particles in the jet such that the secondary subjet lies directly below the origin.
    \item \textbf{Reflect} all particles about the axis $\phi=0$ such that the third subjet has $\phi>0$. If there is no third subjet, reflect such that the $p_T$-weighted sum of $\phi$ over all particles in the jet is positive.
\end{enumerate}
The normalisation step serves to partially reduce the dependence on the $p_T$ of the jet. 
Although the radial scale of the event still varies with the transverse momentum, we do not 
remove this dependence (through zooming~\cite{unimelb} or a similar procedure) since the 
angular scale of the radiation pattern is correlated with the jet type whereas the jet $p_T$ 
itself is not. The remaining steps remove redundancies arising from spatial symmetries, namely 
the location and orientation of the event in the detector.

\subsection{Model details and performance}\label{sec:models}

To evaluate the performance of the equivariant architecture, we employ six neural network 
models which we call EFN, PFN(-ID), EV-EFN and EV-PFN(-ID). The details of each model are 
given below. All nodes in the networks (except for those in an output layer) use the ReLU 
activation\:\cite{relu} and all weights and biases are initialised by He-uniform parameter 
initialisation\:\cite{he-uni}. The Adam optimisation algorithm\:\cite{adam} is used to minimise 
the binary cross-entropy loss function with data split into batches of 250. Equivariant layers 
are fixed to a size of $M$=140 (larger than the number of particles in any of jets in the datasets). 
The EFN, PFN and PFN-ID models are implemented using the \texttt{Energyflow} package\footnote{Available at \url{https://energyflow.network}.} 
for \texttt{Python}\:\cite{efn}. The remaining networks, which involve equivariant layers, were 
trained through \texttt{Keras~\textcolor{black}{2.2.4}}\:\cite{keras} with a \texttt{Tensorflow~\textcolor{black}{1.15}} \cite{tensorflow} backend.\footnote{The code used to construct these models can be found at\\ \url{https://github.com/ayo-ore/equivariant-efns}.}

\begin{figure}[!t]
    \centering
    \includegraphics[width=\linewidth]{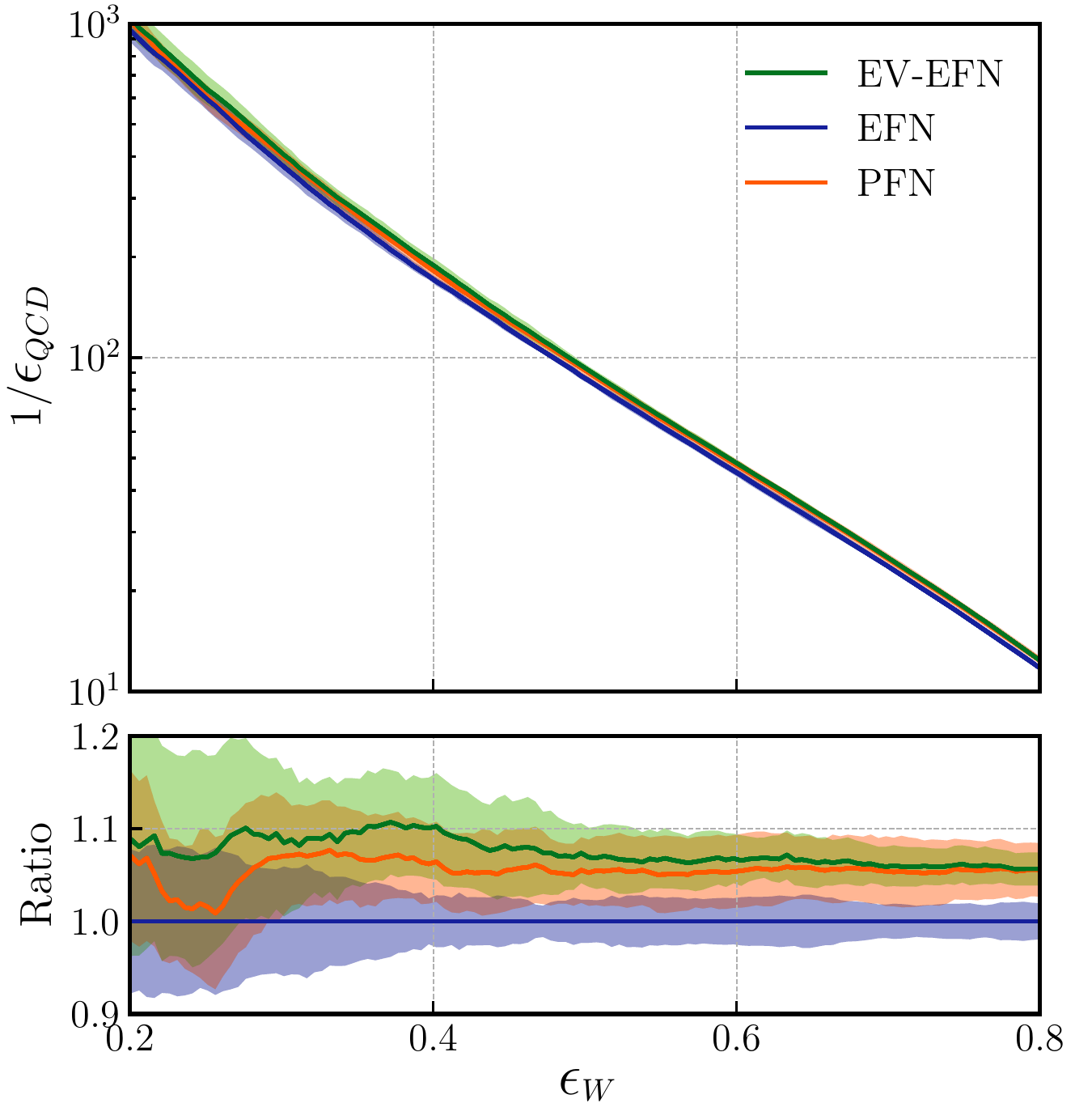}
    \caption{ROC curves averaged over 10 networks for each of the EV-EFN (green), EFN (blue) and PFN (orange) models trained on the $p_T\in[500,550]$ GeV dataset. The bands show one standard deviation. The lower plot shows the ratios relative to the EFN ROC curve.}
    \label{fig:equi-roc}
\end{figure}

We decide on the number of epochs over which to train a model by monitoring the accuracy and loss 
on the validation set during training and stopping when the results begin to diverge from the test 
set. The choices of number and size of layers in the EV models are such that the networks contain \textcolor{black}{a comparable number of parameters as their unaugmented partners. We confirm that equalising the parameter counts (either by increasing or decreasing $\ell$ in the EFN or EV-EFN respectivaly) yields no significant changes to the results presented in this paper.}

\begin{table}[t!]
    \begin{ruledtabular}
    \begin{tabular}{cccc}%{c@{\hspace{3pt}}ccc}
     Model & Jet $p_T$ [GeV] & AUC & $R_{\varepsilon_{W}=0.5}$\\
    \colrule
    \multirow{2}{*}{EFN}       & 500 -- 550 & $0.9339\pm0.0007$ &  $87.4\pm5.7$\\
                               & 250 -- 300 & $0.9092\pm0.0014$ &  $49.1\pm1.6$\\ \hline
    \multirow{2}{*}{EV-EFN }   & 500 -- 550 & $0.9367\pm0.0009$ &  $93.4\pm2.5$\\
                               & 250 -- 300 & $0.9111\pm0.0015$ &  $51.5\pm1.9$\\ \hline
    \multirow{2}{*}{PFN}       & 500 -- 550 & $0.9366\pm0.0012$ &  $93.2\pm4.6$\\
                               & 250 -- 300 & $0.9117\pm0.0011$ &  $51.4\pm1.6$\\ \hline
    \multirow{2}{*}{EV-PFN}    & 500 -- 550 & $0.9333\pm0.0026$ &  $80.1\pm3.8$\\
                               & 250 -- 300 & $0.9083\pm0.0008$ &  $46.2\pm1.6$\\ \hline
    \multirow{2}{*}{PFN-ID}    & 500 -- 550 & $0.9439\pm0.0014$ & $119.0\pm9.0$\\
                               & 250 -- 300 & $0.9180\pm0.0008$ &  $60.0\pm1.8$\\ \hline
    \multirow{2}{*}{EV-PFN-ID} & 500 -- 550 & $0.9402\pm0.0010$ &  $96.0\pm4.0$\\
                               & 250 -- 300 & $0.9152\pm0.0012$ &  $53.5\pm2.6$\\
    \end{tabular}
    \end{ruledtabular}
    \caption{Performance metrics obtained by the models listed in text on both $p_T$ datasets. The background rejection metric $R$ is defined as $1/\varepsilon_{QCD}$ at 50\% signal efficiency. For both the AUC score and $R$, a greater value corresponds to better classification. The reported uncertainties are interquartile ranges of values from 10 separately trained models.}
    \label{tab:equi_performance}
    
\end{table}

The networks 
are defined as follows.

\begin{figure*}
    \centering
    \includegraphics[width=\textwidth]{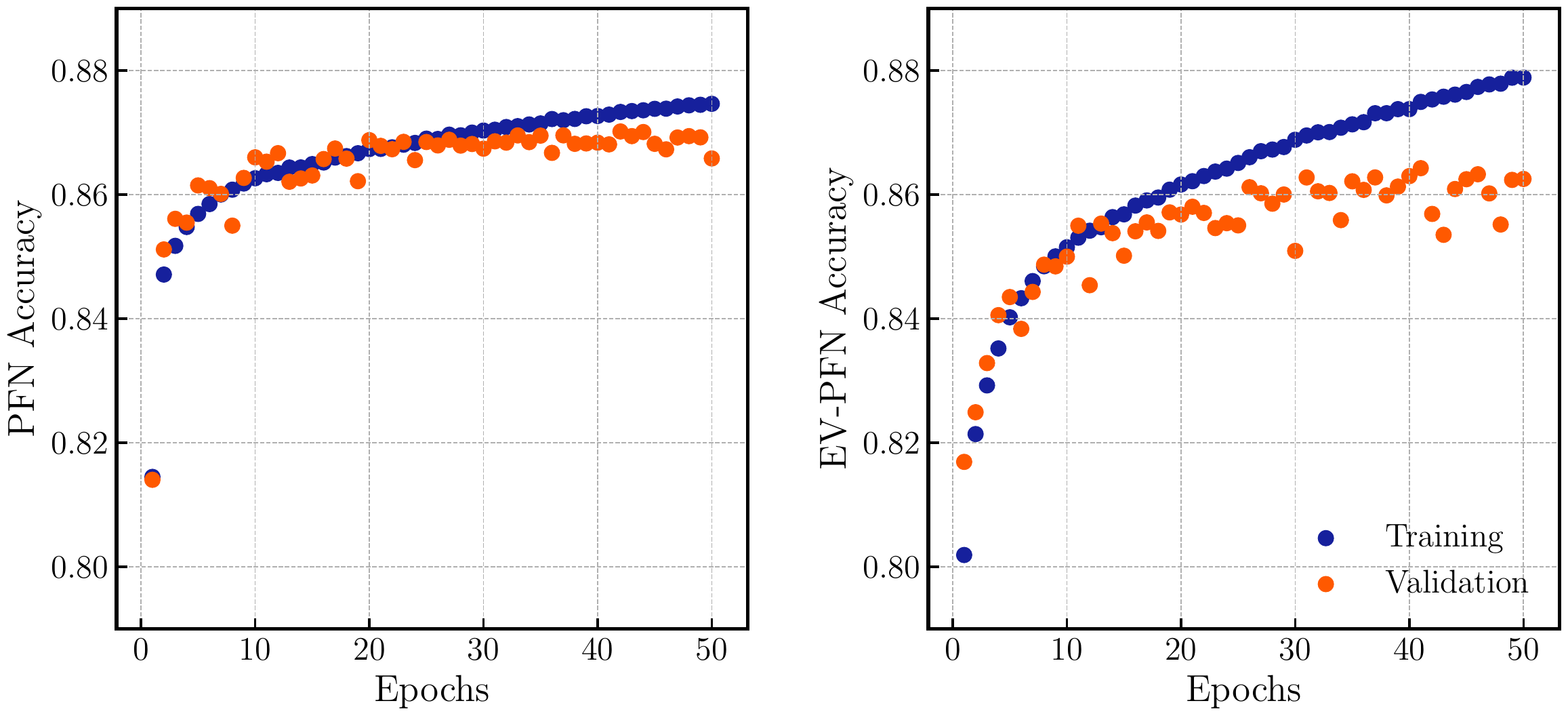}
    \caption{Classification accuracies of the PFN (left) and EV-PFN (right) models on the training and validation subsets of the $p_T\in[500,550]$ GeV data over 50 epochs.}
    \label{overfit}
\end{figure*}

\begin{itemize}[align=left]

\item[\textbf{EFN:}] An $\ell=256$ energy flow network where the network representing the per-particle 
map has two hidden layers of 100 nodes, and the network representing $F$ has three, also with 100 nodes 
each. \textcolor{black}{The model contains 82,358 parameters and} is trained for 30 epochs.

\item[\textbf{PFN and PFN-ID:}] These two networks have the same general construction as the EFN network 
above, however the constituent particles' $p_T$ fractions are given as input to the per-particle map 
instead of weighting the sum over filters. These networks represent the observable decomposition in 
Eq.\,\ref{eq:pfn-obs}, which is not IRC safe in general. In the ID variant, the input feature vector 
contains both momentum and particle identification. The identities are represented by small floats 
beginning at $0$ for an absent particle, then increasing by $0.05$ for each category of constituent: 
$\gamma$, $p$, $\Bar{p}$, $n$, $\Bar{n}$, $\pi^+$, $\pi^-$, $K^+$, $K^-$, $K^0_L$, $e^+$, $e^-$, $\mu^+$, 
$\mu^-$. %A diagram showing the altered per-particle map for both PFNs is shown in Fig.\,$\ref{fig:pid-arch}$.
The models \textcolor{black}{contain 82,458 parameters and} are trained for 25 epochs.

\item[\textbf{EV-EFN:}]  This is an $\ell=128$ EFN containing two IRC safe equivariant layers as 
defined in Eq.\,\ref{eq:irc-equi-def}. We use a smaller latent dimension of $\ell=128$ and reduce 
the depth of the MLP representing the function $F$ such that the total number of parameters in the 
model is comparable to that of the $\ell=256$ EFN. Specifically, the networks for $\Phi$ and $F$ 
both contain two hidden layers of 100 nodes. The equivariant layers have size $M=140$ (corresponding 
to the maximum number of jet constituents) and 100 output channels to match the size of the layers in 
$F$. The model \textcolor{black}{contains 89,330 parameters and} is trained for 30 epochs.

\item[\textbf{EV-PFN and EV-PFN-ID:}] 
These are similar to the $\ell=128$ PFN(-ID) networks, but extended with two equivariant layers in 
the same fashion as the EV-EFN. \textcolor{black}{
The equivariant layers are of the type in Eq.\,\ref{eqvmatmax} (which is not IRC-safe) and max-pooling is used as the projection operation. Additionally, we use 14 dimensional basis vectors to represent each category of constituent particle in the ID variant. Although these choices for PID encoding and projection are different to those in the PFN(-ID) models, we find that they lead to better stability of training. Models using other choices (such as sum-pooling or the use of equivariant layers of the type in Eq.\,\ref{eqvmatsum}) often failed to converge, having AUC scores of 0.5 after training. In the instances where training did converge, results similar to those achieved by the architecture outlined here were observed. The EV-PFN(-ID) models contain 89,430 (90,830) parameters and, due to poor generalisation, are trained for 20 epochs.
}

\end{itemize}

The AUC scores and background rejections, ${R\equiv1/\varepsilon_{QCD}}$, for the models trained on the 
low- and high- $p_T$ datasets are presented in Tab.~\ref{tab:equi_performance}. Comparing the IRC safe models, 
we see that the EV-EFN achieves slightly better classification than the EFN across both measures in both $p_T$ 
ranges. We also show the ROC curves averaged over 10 networks for these models (as well as the PFN) in 
Fig.\,\ref{fig:equi-roc}. The curves here again reflect the classification advantage that the EV-EFN achieves 
over the EFN. This suggests that the addition of equivariant layers aids the network's ability to learn from 
IRC safe information. In fact, the improvement obtained by the EV-EFN is greater than the improvement seen from 
the PFN model, which is not IRC safe and uses a latent space of twice the dimensionality. However, the results in 
Tab. \ref{tab:equi_performance} show that the EV-PFN does not discriminate better than the PFN model. In fact, on 
both datasets it performs worse than even the IRC safe EFN across both metrics. The result for the EV-PFN-ID model 
is not as poor -- although it does not match the classification power of the PFN-ID networks, it does at least 
outperform the PFN. 

A major issue that the EV-PFN(-ID) networks face is \textcolor{black}{poor generalisation}. To illustrate this, in Fig.\,\ref{overfit} we 
plot the classification accuracy of the PFN and EV-PFN models on the training and validation datasets over 50 
epochs. The test accuracy for the EV-PFN diverges from the validation accuracy after just 15 epochs of training, 
whereas this behaviour is not seen to such a strong degree for the PFN despite both models containing \textcolor{black}{approximately} the same 
number of parameters.

The plots reveal that the EV-PFN does indeed classify jets from the training set more accurately than the PFN, 
however its poor generalisation means that this performance is not transferred to test results. This could possibly be addressed by regularisation techniques. One such option is to apply 
\emph{dropout} -- a method wherein nodes in a layer are randomly selected to be switched off during 
training\:\cite{dropout}. We investigate application of dropout on layers in the network for $F$, but find 
that for a variety of dropout rates, only the performance on the training set is affected. The same result is observed 
for the ID variants. In the interest of avoiding potentially unfruitful trial and error, we leave \textcolor{black}{fine-tuning} 
of the EV-PFN(-ID) models to future work.

\subsection{Dependence on jet mass}
\begin{figure*}[t!]
    \centering
    \includegraphics[width=\textwidth]{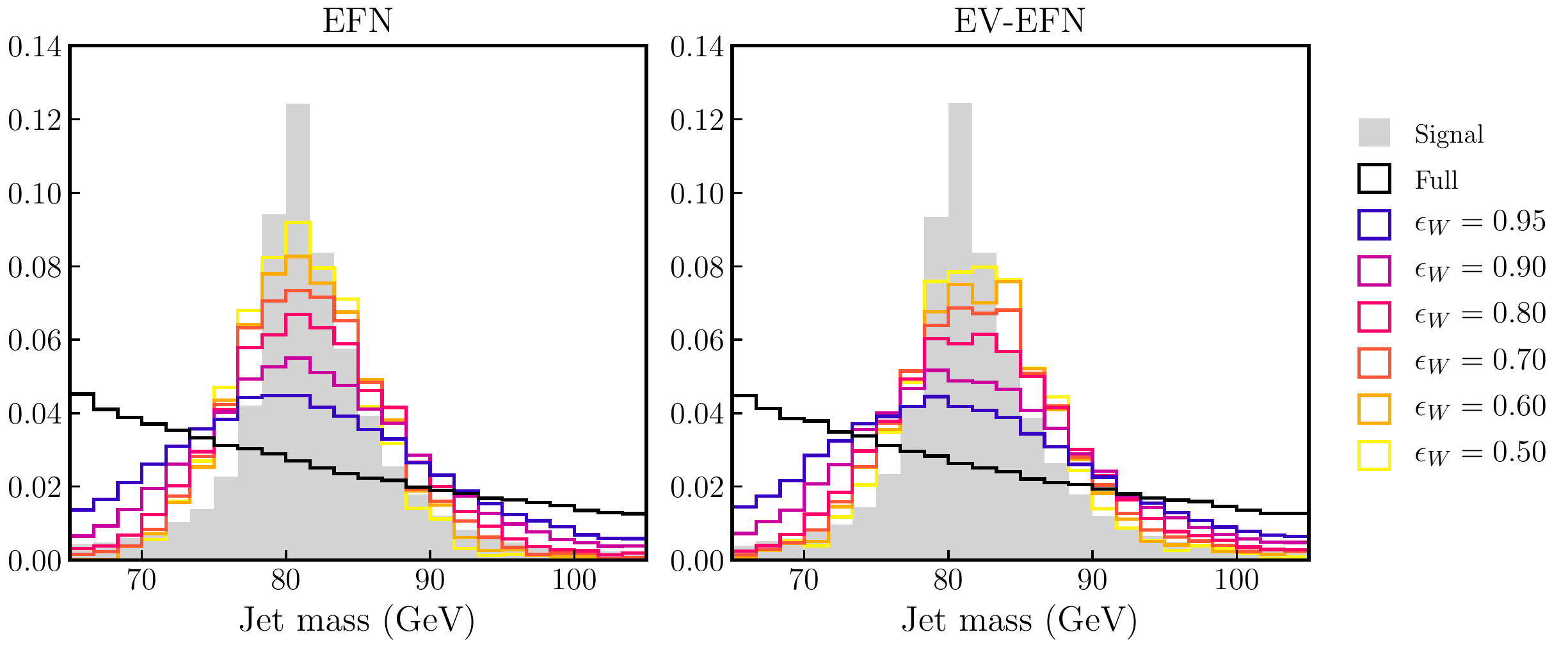}
    \caption{Jet mass histograms (normalised) for the $p_T\in[500,550]$ GeV QCD jets that are tagged as $W$'s by successively harder cuts on the EFN (left) or EV-EFN (right) output. The distribution for $W$ jets is shown in grey.}
    \label{fig:network-sculpting}
\end{figure*}

In this section, we compare the effect of the EV-EFN and EFN on the QCD jet mass distribution. We follow closely 
the approach taken in Ref.\,\cite{mass-agnostic}. An unrestricted neural network will make use of all 
available information, jet mass included. Consequently, the output of such networks will be correlated 
with the jet mass, leading to a sculpting effect of the jet mass distribution. This is not a desirable 
feature, since background modeling is often done by extrapolating from a region containing no signal, 
relying on an approximately flat background. A background distribution that is smooth and featureless 
is associated with less systematic uncertainty. There exists a trade-off between the raw discrimination 
power a neural network offers and the degree to which it reshapes the background jet mass distribution. 
For this reason, jet observables that are strongly correlated with the jet mass are less valuable for 
experimental use compared to those that are uncorrelated. This problem has been studied extensively from 
both analytic~\cite{DDT, analytic-decorrelation} and machine learning perspectives~\cite{data-planing, pca, 
adversarial, uboost, DisCo, MUST,ATLAS:2018ibz, mass-agnostic, Dorigo:2020ldg}.

Here we compare results first for the standard training procedure described above and then including the 
application of a method for decorrelating network output from the jet mass. To compare the two models, 
we train 10 EFNs and 10 EV-EFNs on the $p_T\in[500,550]$ GeV dataset, split into training (75\%) and 
testing (25\%) subsets. By increasing the size of the testing set, we reduce the impact of statistical 
fluctuations on the results. This is important since, at low signal efficiencies, a very small number 
of QCD jets are tagged. In Fig.\,\ref{fig:network-sculpting} we visualise the sculpting effect of the 
EFN and EV-EFN models on the training set by showing the jet mass distributions of QCD jets that are 
tagged as signal by successively tighter cuts the network output. Both models strongly sculpt the QCD 
distribution to the $W$ mass peak.  In order to quantify the degree of sculpting we make use of the 
Hellinger distance, defined as

\begin{equation}
\newcommand*\bigstrut{\vrule height 1.3\baselineskip depth.3\baselineskip width 0pt\relax}
    H(p,q) = \sqrt{\bigstrut 1-\sum_i\sqrt{p(i)q(i)}}\ ,
\end{equation}
where $p$ and $q$ are discrete probability distributions. Hellinger distances are normalised between 
0 and 1, with a distance of $0$ corresponding to $p$ and $q$ being identical histograms. For each 
trained model, the Hellinger distances between the jet mass distributions of tagged QCD jets and the 
original are calculated (using histograms of 25 bins) for a range of cuts on network output. We then 
average the distances over network instances.

\begin{figure*}[tp!]
    \centering
    \includegraphics[width=\textwidth]{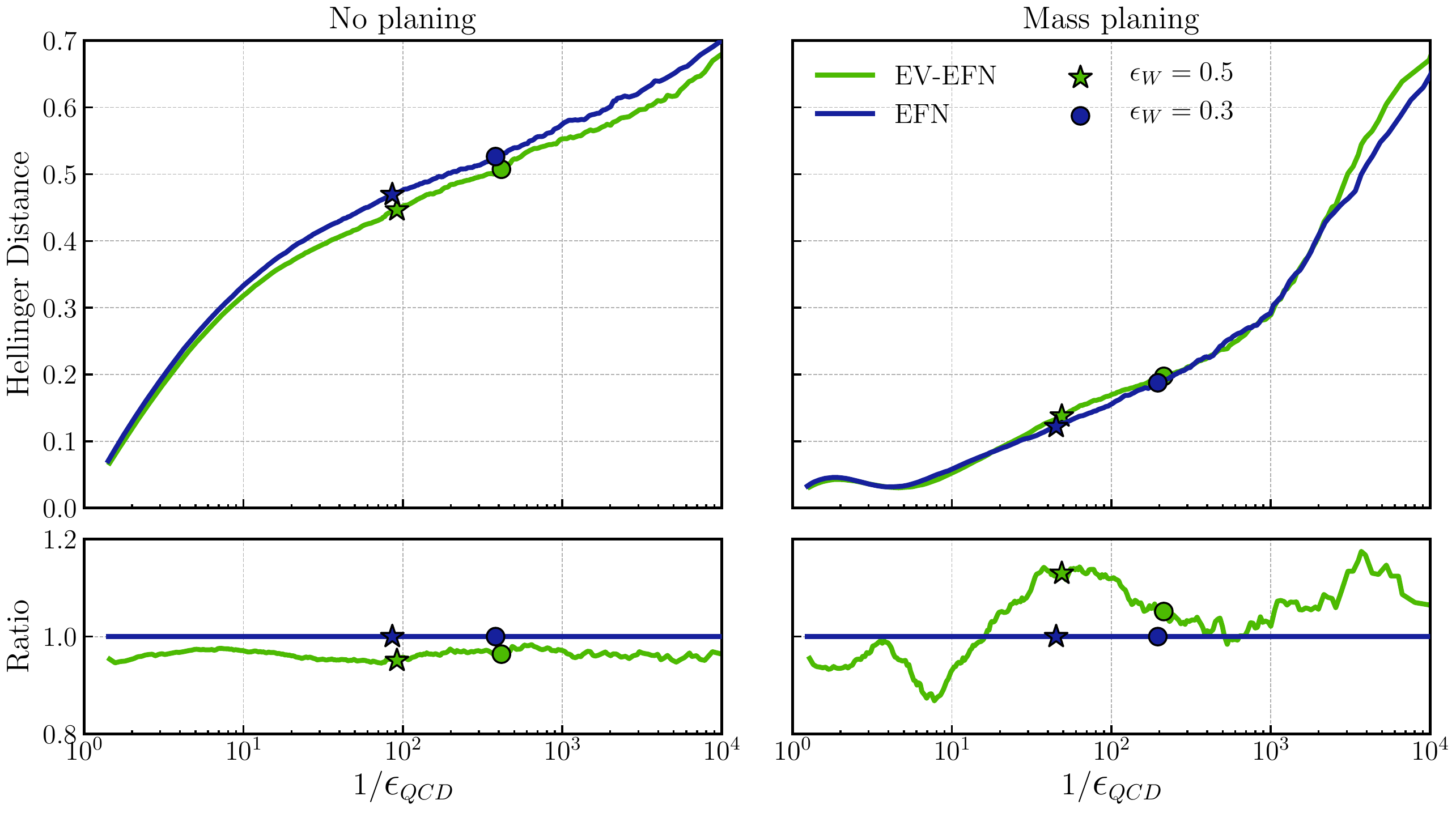}
    \caption{Hellinger distances between the tagged QCD jet mass distribution and the full distribution as a function of the background rejection achieved by cutting on the output of an EFN or EV-EFN. Results are shown for the standard training procedure (left) and with the inclusion of mass-planing (right). We indicate with markers the positions along the curves corresponding to fixed signal efficiencies of $\varepsilon_W=0.3$ or $\varepsilon_W=0.5$. The ratios to the EFN-line are shown beneath each plot.}
    \label{fig:hellinger}
\end{figure*}

A plot of these average distances as a function of the network background rejection for the two models 
is shown on the left of Fig.\,\ref{fig:hellinger}. In these terms a better classifier corresponds to a 
curve that is closer to the bottom-right corner, reaching greater background rejection at smaller 
Hellinger distance. The general behaviour of the two models is very similar, however the EV-EFNs achieve 
smaller Hellinger distances across all cuts on the network output. The greatest difference is seen at a 
background rejection of $1/\varepsilon_{QCD}\sim90$ (corresponding to a signal efficiency of 
$\varepsilon_W=0.5$), where they improve upon the EFNs by approximately 5\%.

While the EV-EFNs have a smaller effect on the background distribution compared to the EFNs, both 
models sculpt to such a great degree that this improvement can be considered somewhat inconsequential. 
Having said this, a more meaningful result can be obtained by investigating the networks' sculpting after 
decorrelating the network output from the jet mass. Ref.\,\cite{mass-agnostic} found that for neural 
network taggers, techniques that achieve this through adjustments to the input data provide decorrelation 
almost on par with those that alter the training procedure, while having significantly lower computational 
cost. For this reason, we compare the response of EFNs and EV-EFNs to \emph{mass-planing} -- a particular 
case of \emph{data planing} \cite{data-planing,jet-images}. A dataset can be planed in some observable, 
$\rho$, by assigning each instance of the training data $x_i$ with a weight $w$ according to
\begin{equation}
    \left[w(x_i)\right]^{-1} = \frac{d\sigma_x}{d\rho}\bigg|_{\rho=\rho(x_i)},
\end{equation}
where $\frac{d\sigma_x}{d\rho}$ is the differential cross-section for jets in $x$. In practice, this 
simply corresponds to inverting the histogram for $\rho$ at the bin in which $x_i$ lies. After such a 
weighting, the data has a uniform distribution in the planed observable.

To implement mass-planing in the networks' training procedure, we calculate weights from the $W$ and 
QCD mass distributions histogrammed into 100 bins. The weights are used to give the relative importance 
of a jet's contribution to the loss function during training. The Hellinger distances are then calculated 
and averaged in the same fashion as for the un-planed case. On the right side of Fig.\,\ref{fig:hellinger}, 
we again present a plot of the calculated distances against background rejection for the newly trained 
networks. Interestingly, the advantage that the EV-EFN displayed without planing is no longer apparent. 
In fact, the standard EFN exhibits less background sculpting across most cuts on the network output. 
One should therefore be cautious about interpreting the absolute performance of some given taggers 
without considering decorrelation, as this may change the order of the rankings. However, the EV-EFN 
still achieves greater background rejection at fixed signal efficiency. This implies that in a bump 
hunt scenario, neither model has a clear advantage in terms of achieved significance. 

% Section 5
\section{Conclusions}

The development of new and increasingly accurate jet taggers remains a high priority for the collider 
physics community. The use of techniques based on machine-learning methods has provided a number of 
advances in this area in the past few years. Recent research has explored incorporating permutation 
invariance of the constituent particle momenta into the network architecture via the Deep Sets formalism, 
leading to Energy Flow Networks (EFNs). However, Deep Sets also allows for the possibility of permutation 
equivariance, which is what we have explored in this paper.

We have successfully implemented permutation-equivariant neural network layers into the Energy Flow 
Network architecture. Furthermore, we identified a specific construction, the EV-EFN,  that maintains 
the infrared and collinear safety of the learned jet observable. When trained to discriminate $W$ from 
QCD jets in simulated samples in two boosted $p_T$ windows these EV-EFNs exhibited a marginal improvement 
in performance over permutation-invariant EFNs. 
The performance of the EV-EFN was equal to that achieved by a Particle Flow Network, which is not 
restricted by IRC safety. \textcolor{black}{Some issues were observed when extending PFNs by equivariant layers. Specifically, certain choices of equivariant operation or projection prevented the model from converging during training. Further, models that did converge generalised poorly to the testing sample. These issues could possibly be remedied by alternative constructions or regularisation techniques, however we leave that task for future work.}

Finally, we compared the action of the networks on the QCD jet mass distribution before and after 
applying mass planing. The planing decorrelates the tagger performance from the jet mass, at some 
cost in performance. However, it also changes the relative performances of the EFN and EV-EFN networks. 
While the EV-EFN initially had slightly less impact on the background distribution compared to the EFN, 
this advantage was lost after mass planing the input jets. Consequently it would be interesting to apply 
recent ideas such as Distance Correlation\,\cite{DisCo}, adversarial networks\,\cite{adversarial}, Mass 
Unspecific\,\cite{MUST} (or Agnostic\,\cite{mass-agnostic}) Supervised Tagging to see if networks with 
equivariant layers continue to outperform invariant ones. 

A number of different future directions present themselves. Future work could investigate the use of 
other types of equivariant layers and their impact on IRC safety including those which are equivariant 
to continuous symmetries of the Lorentz Group as in Ref.\,\cite{Bogatskiy:2020tje} and similar to 
Ref.\,\cite{Butter:2017cot}. It may also be interesting to attempt to translate the EV-EFN into a 
low-dimensional human-interpretable space using the method outlined in Ref.\,\cite{human-read}. This 
could potentially illuminate the effect that equivariant layers have on the  representation of IRC 
safe jet observables compared to the original EFN. 

% Acknowledgements
\section*{Acknowledgements}

We thank John Gargalionis for useful comments and discussions. This work was supported in part by 
the Australian Research Council and the Australian Government Research Training Program Scholarship initiative.

\appendix*
\textcolor{black}{
\section{Impact of equivariance}\label{app:equi-test}
As mentioned in Section\,\ref{sec:irc-safety}, under certain restrictions to the equivariant layers, the architecture of an EV-EFN reduces to that of a standard EFN. These restrictions include taking one of $\sigma\equiv id$, $\Gamma=0$ or $\Lambda=0$. The first case corresponds to rescaling the output of $\Phi$ as shown in Eq.\,\ref{eq:id-act}. The other two cases allow the equivariant layers to be absorbed as additional layers in $\Phi$ or $F$ respectively.}

\textcolor{black}{
Here we evaluate the performance of EV-EFNs (with architecture as described in Section\,\ref{sec:models}) trained under each of these restrictions. The results are presented in Tab.\,\ref{tab:equi-test} and show that the performance of the EV-EFN degrades in all cases. This demonstrates that the improvement of the EV-EFN over the EFN is not simply due to the model's increased depth.
}
\begin{table}[h!]
    \textcolor{black}{
    \centering
    \begin{ruledtabular}
    \begin{tabular}{ccc}%{c@{\hspace{3pt}}ccc}
     Model & AUC & $R_{\varepsilon_{W}=0.5}$\\
    \colrule
    \phantom{$\big|_{\sigma\equiv id}$}EV-EFN$\big|_{\sigma\equiv id}$   & $0.9320\pm0.0009$ & $83.5\pm5.2$\\
    \phantom{$\big|_{\Gamma=0}$}EV-EFN$\big|_{\Gamma=0}$   & $0.9327\pm0.0013$ & $85.0\pm1.9$\\
    \phantom{$\big|_{\Lambda=0}$}EV-EFN$\big|_{\Lambda=0}$   & $0.9337\pm0.0008$ & $87.6\pm2.4$\\
    EFN      & $0.9339\pm0.0007$ &  $87.4\pm5.7$\\
    EV-EFN   & $0.9367\pm0.0009$ &  $93.4\pm2.5$
    \end{tabular}
    \end{ruledtabular}
    }
    \caption{\textcolor{black}{Classification metrics on the $p_T\in[500,550]$ GeV dataset achieved by EV-EFNs with restricted equivariant layers in comparison to the full EV-EFN and the base EFN.}}
    \label{tab:equi-test}
\end{table}


% \bibliographystyle{aps-revtex}
% \bibliographystyle{apsrev4-1}
%\bibliography{references}

\end{document}